\documentclass{article}
\usepackage[table,xcdraw]{xcolor}
\usepackage{spconf,amsmath,graphicx}

\usepackage{mathtools}

\usepackage{caption}
\usepackage{subcaption}

\DeclarePairedDelimiter\abs{\lvert}{\rvert}

\title{Text-free non-parallel many-to-many voice conversion\\using normalising flows}

\name{\begin{tabular}{c}Thomas Merritt$^1$, Abdelhamid Ezzerg$^1$, Piotr Biliński$^1$, Magdalena Proszewska$^2$\sthanks{Work performed during an internship at Amazon.}, \\
Kamil Pokora$^1$, Roberto Barra-Chicote$^1$, Daniel Korzekwa$^1$\end{tabular}}
\address{\begin{tabular}{c}$^1$ Amazon Alexa, $^2$ Jagiellonian University, Poland \\
\{thommer, ezzerg, bilipiot, kamipoko, rchicote, korzekwa\}@amazon.com
\end{tabular}}

\begin{document}
\ninept
\maketitle
\begin{abstract}
Non-parallel voice conversion (VC) is typically achieved using lossy representations of the source speech. However, ensuring only speaker identity information is dropped whilst all other information from the source speech is retained is a large challenge. This is particularly challenging in the scenario where at inference-time we have no knowledge of the text being read, i.e., text-free VC. 
To mitigate this, we investigate information-preserving VC approaches.

Normalising flows have gained attention for text-to-speech synthesis, however have been under-explored for VC. Flows utilize invertible functions to learn the likelihood of the data, thus provide a lossless encoding of speech.
We investigate normalising flows for VC in both text-conditioned and text-free scenarios. Furthermore, for text-free VC we compare pre-trained and jointly-learnt priors.
Flow-based VC evaluations show no degradation between text-free and text-conditioned VC, resulting in improvements over the state-of-the-art. Also, joint-training of the prior is found to negatively impact text-free VC quality.

\end{abstract}
\vspace{-2mm}
\section{Introduction}
\label{sec:intro}
\vspace{-2mm}
Voice conversion (VC) is the process of taking speech from a source speaker and changing the perceived speaker identity to match that of a target speaker \cite{mohammadi2017overview,sisman2020overview}. The result is that we are able to change source speech to sound as though it was read by the target speaker. There are two major families of training paradigms for VC: parallel and non-parallel \cite{lorenzo2018voice}. Parallel VC relies on the same text being read by source and target speakers, therefore allowing to train the model directly on the VC task (i.e., having ground truth paired source and target speech for training). However, such datasets are expensive to record and challenging to enrol speakers. Therefore, more commonly non-parallel methods are applied, which have no such data requirements. Non-parallel VC approaches allow for much easier enrolment of speakers as the only requirement is transcribed speech. This work focuses on the non-parallel VC scenario.

Typical non-parallel VC approaches in literature learn lossy representations of the source speech to achieve successful conversion \cite{qian2019autovc,karlapati2020copycat,van2017neural,polyak2020tts,qian2020unsupervised,wang2021adversarially}. These models are encouraged to drop information that we wish to change (i.e., speaker identity) whilst retaining all other aspects of the source speech (i.e., the text being read and prosody). This makes the approach challenging to train to ensure that the balance between retaining and losing information from the source speech is achieved. Therefore, in this investigation we explore information-preserving approaches to VC.

Whilst there has been numerous previous works investigating the application of normalising flows for speech, these have typically focused on text-to-speech (TTS) \cite{kim2020glowtts,miao2020flow,valle2020flowtron,casanova2021sc} or neural vocoding \cite{oord2018parallel,prenger2019waveglow} rather than VC \cite{serra2019blow}. Normalising flows, learn an invertible mapping of the input data to a latent vector \cite{kingma2018glow,ho2019flow++,kobyzev2020normalizing}. During training the latent vector is trained to maximise the likelihood of a prior distribution. 
Typically, conditioning features are passed to the flow steps to help the model to better maximise this likelihood \cite{serra2019blow,miao2020flow,valle2020flowtron,casanova2021sc}. 
The motivation for this is that the network can learn to remove the contributions of these conditioning features from the latent vector, hence removing the level of noise to the prior.  
The invertible nature of the flow ensures that the network is information-preserving, meaning that the latent vector is a lossless representation of the input data. 
Given the challenges described above with designing models using lossy speech representations for VC, the application of normalised flows for VC is of clear interest. 
Provided that the prior distribution is disentangled from the speech attribute we wish to convert, the network should be capable of converting the speech attribute from the source speech, as long as it has successfully learnt its contribution. This allows for a paradigm to perform disentangled conversion of multiple speech attributes (for example speaker identity, accent and speaking style). Here the focus of the investigation is conversion of speaker identity (i.e., VC), however we leave exploration of further speech attributes within this paradigm as future work.

A similar application of normalised flow for VC was explored in \cite{serra2019blow}. However, this was conducted on raw audio input, 
a noisy representation of speech, resulting in conversion being more challenging. 
Mel-spectrograms are a more stable speech representation, hence is used in the majority of recent work in TTS and VC literature \cite{shen2018natural,karlapati2020copycat}.
Therefore, this investigation focuses on mel-spectrograms instead of raw waveforms. 
In addition, the investigation in \cite{serra2019blow} did not explore the use of alternative priors. These are explored in this investigation. Using Glow-TTS for voice conversion is mentioned in \cite{kim2020glowtts,casanova2021sc} however was not formally evaluated.

The contributions of this paper are: 1) formal investigation into normalising flows for the problems of both text-conditioned and text-free VC, resulting in significant improvements over state-of-the-art text-free VC, 
2) for text-free VC 
using 
normalising flows we investigate whether it is better to use a pre-trained ``expert'' prior or whether joint-training of prior and flow are beneficial. 
This finds that joint-training of prior and flow negatively impacts the convergence of the model, 
indicating that improvements to training schedule will likely improve the state-of-the-art for both TTS and VC.

\vspace{-3mm}
\section{Proposed flow-based VC approaches}
\label{sec:flows}
\vspace{-3mm}

\begin{figure*}
\centering
        \begin{subfigure}[b]{0.4\textwidth}
                \centering
                \includegraphics[width=\linewidth]{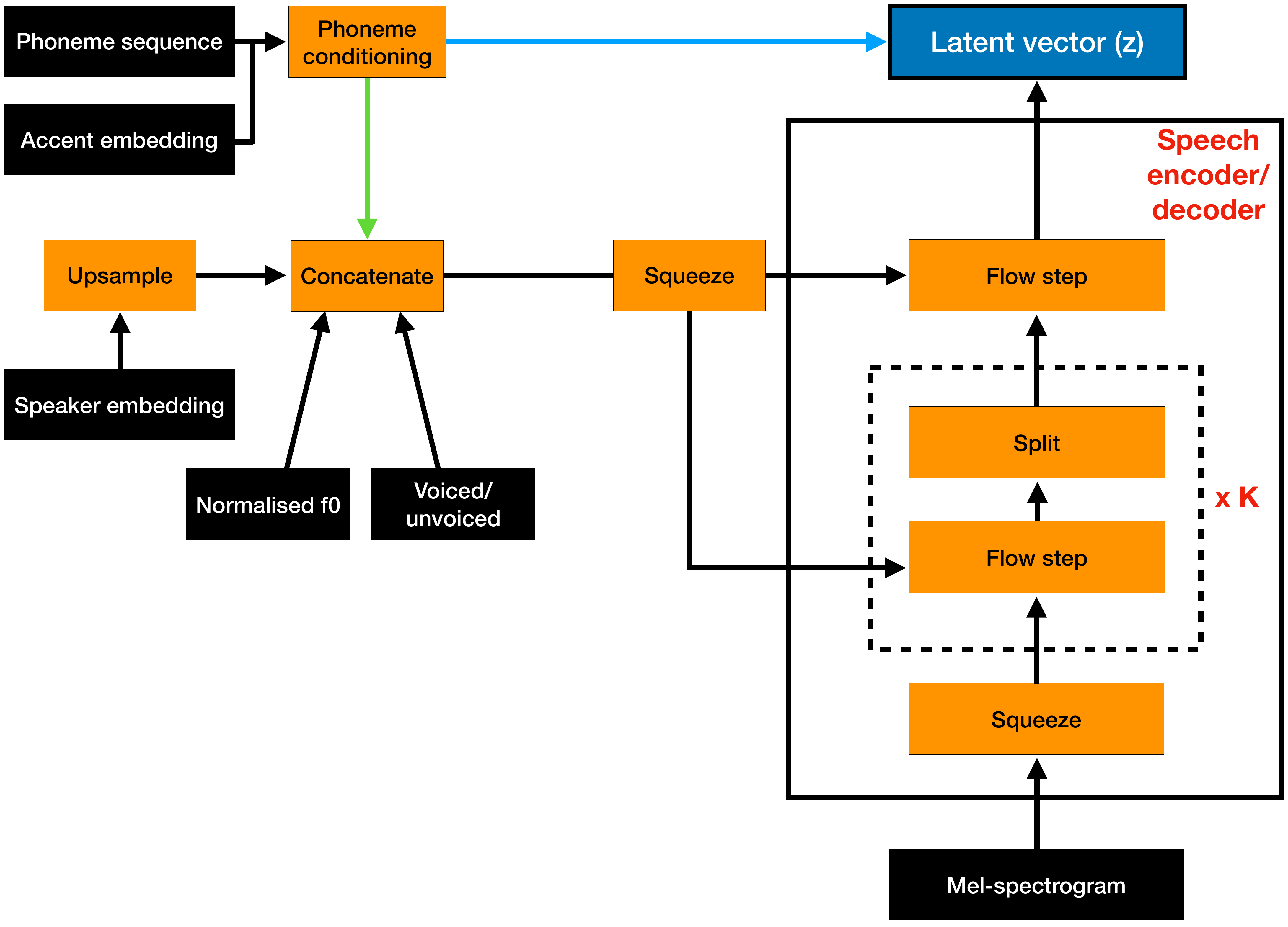}
                \caption{}
                \label{fig:flow_overall}
        \end{subfigure}
        \hspace{-12mm}
        \begin{subfigure}[b]{0.36\textwidth}
                \centering
                \includegraphics[width=0.6\linewidth]{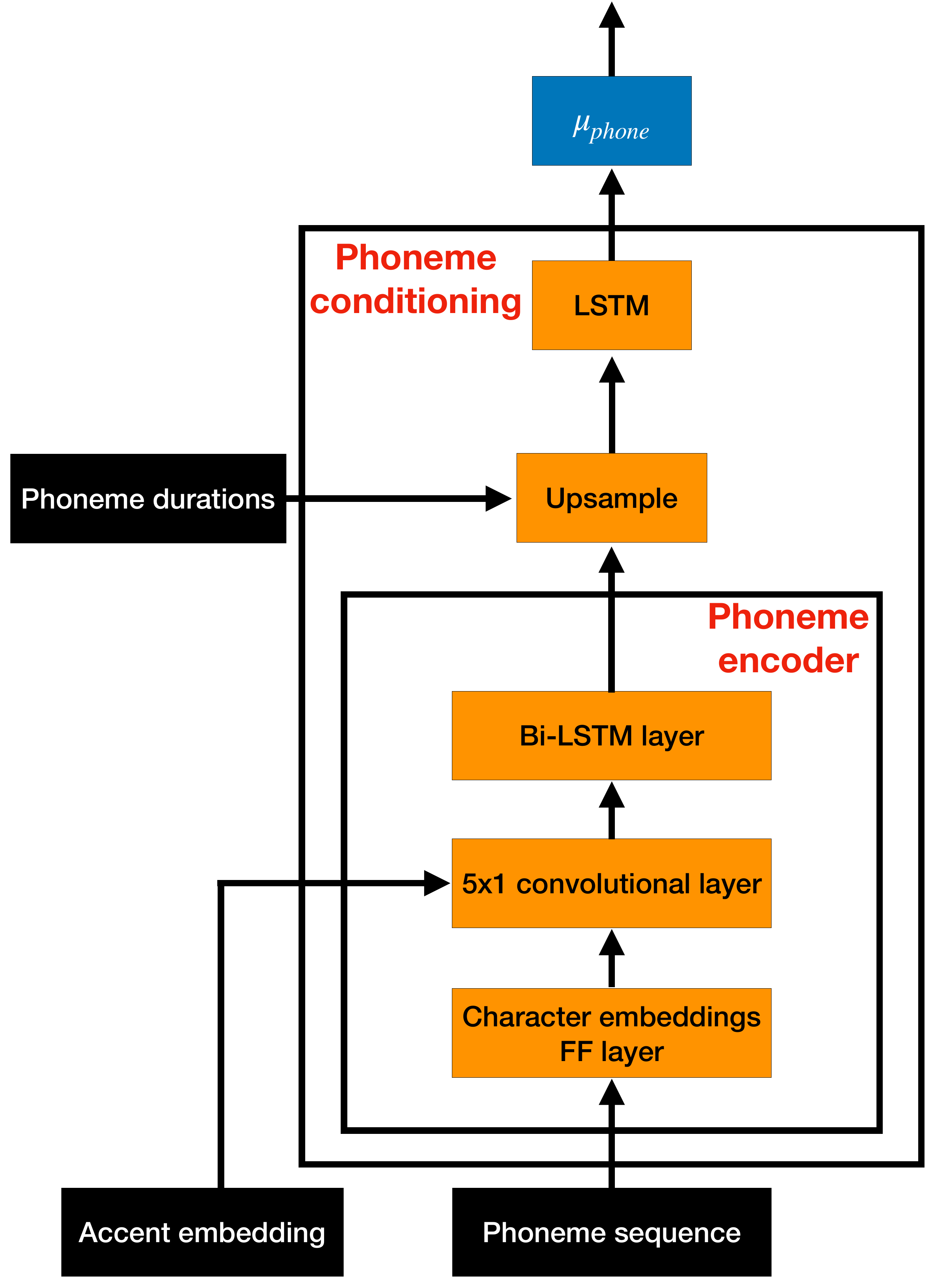}
                \caption{}
                \label{fig:phone_cnd}
        \end{subfigure}
        \hspace{-13mm}
        \begin{subfigure}[b]{0.19\textwidth}
                \centering
                \raisebox{6mm}{\includegraphics[width=\linewidth]{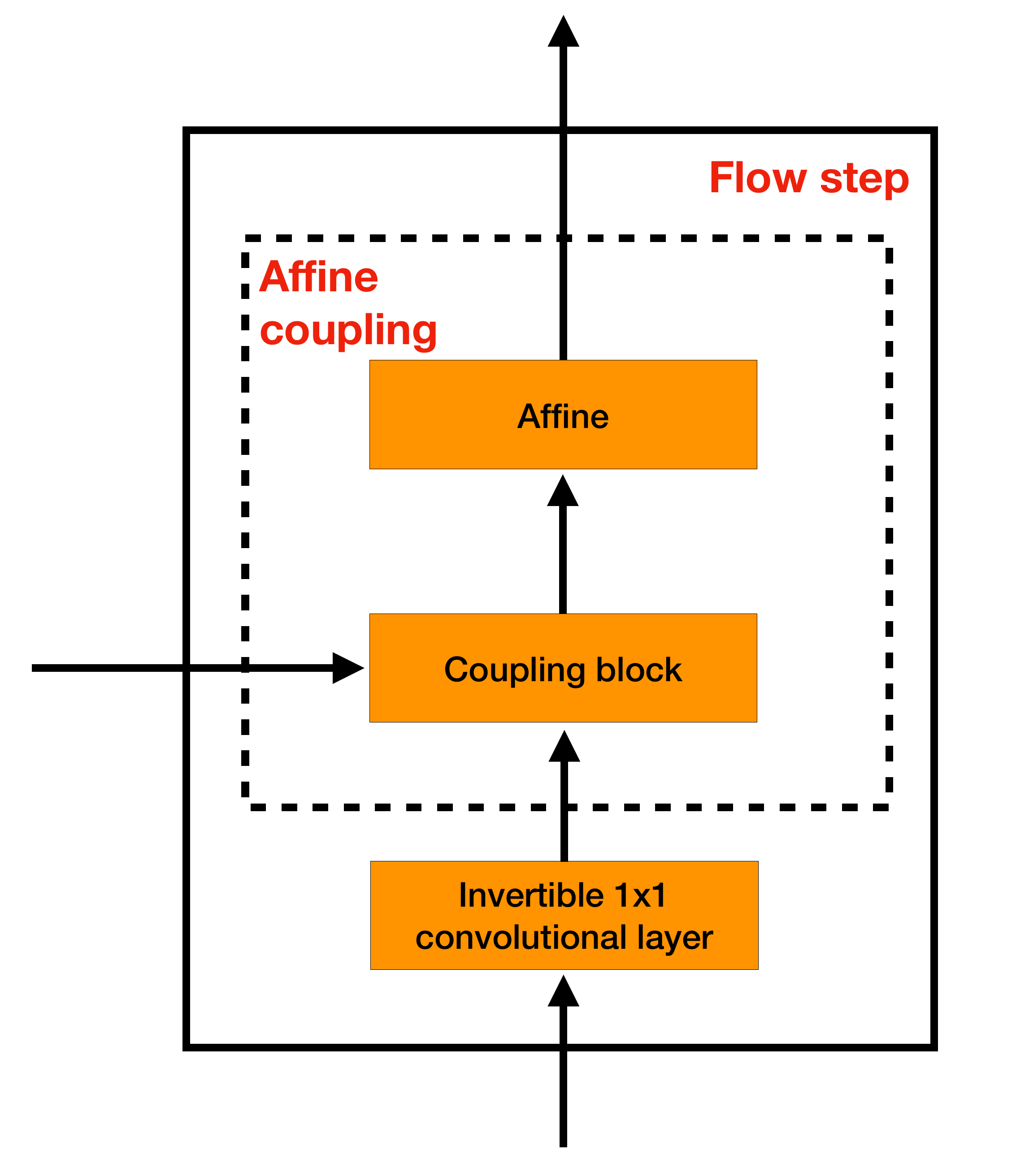}}
                \caption{}
                \label{fig:flow_step}
        \end{subfigure}
         \hspace{-3mm}
        \begin{subfigure}[b]{0.19\textwidth}
                \centering
                \raisebox{4mm}{\includegraphics[width=\linewidth]{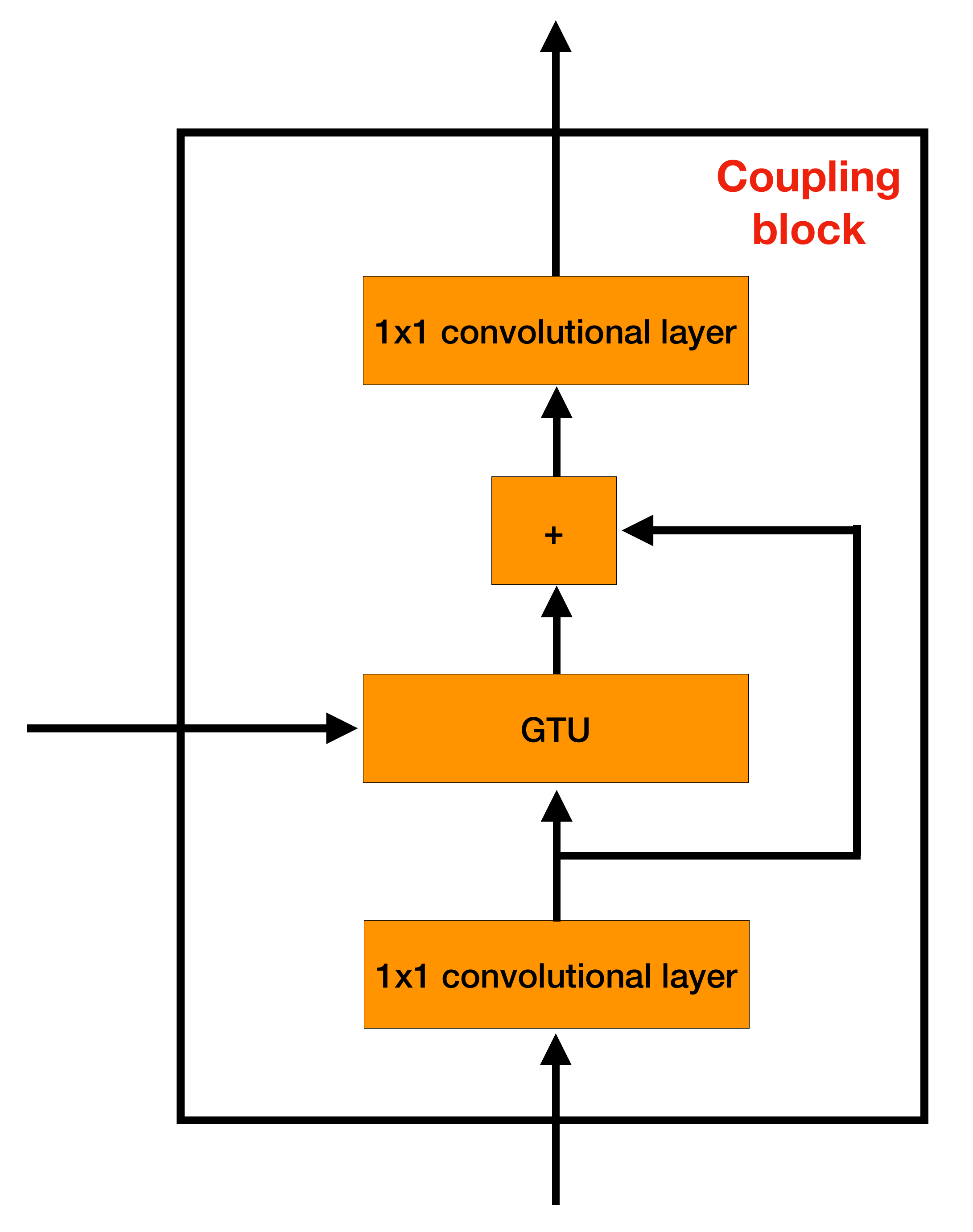}}
                \caption{}
                \label{fig:flow_coupling}
        \end{subfigure}
        \vspace{-2mm}
        \caption{Overview of our VC approach using normalised flows (a) followed by phoneme conditioning component (b), Flow step (c) and Coupling block (d). Green line denotes phoneme conditioning for text-conditioned VC models. Blue line denotes the phoneme-based prior used for text-free VC models.}\label{fig:flowvc}
        \vspace{-6mm}
\end{figure*}

Flow-TTS has demonstrated state-of-the-art quality for TTS \cite{miao2020flow} so was selected as the basis for the model topology of the proposed VC approach. Two variants of VC with this topology are investigated: 1) text-conditioned VC, where text information is provided to the model at both training-time and inference-time, 2) text-free VC, where text information is only required at training-time.

\vspace{-2mm}
\subsection{Text-conditioned VC: fixed prior}
\label{subsec:txt_cnd}
\vspace{-2mm}
Text-conditioned VC is most similar to the original Flow-TTS implementation \cite{miao2020flow}. 
The model topology is shown in Figure \ref{fig:flowvc}, with arrows arranged according to training-time. 
During training-time the model `encodes' input mel-spectrograms ($x$) into the latent vector $z$, as shown in Equation \ref{eq:encode}:

\vspace{-4mm}
\begin{equation}
z = f(x; ph_{source}, spk_{source}, f0_{source}, vuv_{source})
\label{eq:encode}
\end{equation}

\noindent where $ph_{source}$ denotes phoneme conditioning coming from the phoneme encoder (the green arrow in Figure \ref{fig:flow_overall}, $ph_{source}=\mu_{phone}$ 
from Figure \ref{fig:phone_cnd} for the source phoneme sequence), 
$spk_{source}$ is a pre-trained utterance-level speaker embedding~\cite{spk_acc_embeddings}, $f0_{source}$ is normalised interpolated log-f0 and $vuv_{source}$ is a binary `voiced or unvoiced' flag denoting whether a frame is voiced or unvoiced. 
Sentence-level mean normalisation is applied to interpolated f0 
to disentangle speaker identity (i.e., relating to the speaker's average f0) from sentence prosody, thus separating f0 conditioning from speaker embedding conditioning.
Providing this conditioning information allows the network to learn to remove these factors from $z$ in order to better maximise the likelihood of the prior distribution ($c$) during training: 

\vspace{-4mm}
\begin{equation}
log P_{X}(x|c) = log P_{Z}(z|c) + log \abs*{det \frac{\partial z}{ \partial x}}
\label{eq:likelihood_loss}
\vspace{-4mm}
\end{equation}
where:
\begin{equation}
c \sim N(\mu_{prior},\sigma_{prior})
\end{equation}

For text-conditioned VC a uniform prior of $\mu_{prior}=0, \sigma_{prior}=1$ is applied for all frames regardless of the speech content. To perform VC, source speech is encoded according to Equation \ref{eq:encode}, and then converted to the target speaker using Equation \ref{eq:decode}:

\vspace{-4mm}
\begin{equation}
x_{converted} = f^{-1}(z; ph_{source}, spk_{target}, f0_{source}, vuv_{source})
\label{eq:decode}
\end{equation}
\noindent where $spk_{target}$ denotes the speaker embedding of the target speaker.
This is similar to the conversion approach 
in \cite{serra2019blow}.

In previous work on Flow-TTS and Glow-TTS, the phoneme alignment is learnt during training using either attention \cite{miao2020flow} or monotonic alignment search (MAS) \cite{kim2020glowtts}.  
However, here pre-trained phoneme alignments are instead used to label the data, as was done in \cite{shah2021non} for attention-free TTS, simplifying the training process.

\vspace{-2mm}
\subsection{Text-free VC: context-dependent prior}
\label{subsec:txt_free}
\vspace{-2mm}
This approach is more similar to that of Glow-TTS \cite{kim2020glowtts,casanova2021sc}. 
The model is shown in Figure \ref{fig:flowvc}, 
with the blue line in Figure \ref{fig:flow_overall} used instead of the green line. 
Unlike the text-conditioned approach, text-free VC does not pass information relating to the phonetic context to the flow steps. Instead, a phoneme-dependent prior distribution ($c$) is used in Equation \ref{eq:likelihood_loss} to train the speech encoder. 
As a result, the model learns to represent phonetic context within $z$, meaning no prior knowledge of the text being read is required to perform VC. 
Therefore Equations \ref{eq:encode} and \ref{eq:decode} for this scenario become:

\vspace{-2mm}
\begin{equation}
z = f(x; spk_{source}, f0_{source}, vuv_{source})
\label{eq:txt_free_encode}
\end{equation}
and
\vspace{-2mm}
\begin{equation}
x_{converted} = f^{-1}(z; spk_{target}, f0_{source}, vuv_{source})
\vspace{-2mm}
\label{eq:txt_free_decode}
\end{equation}

We explore two options to train the text-free VC approach: 1) use a pre-trained phonetic prior, 2) jointly-train the prior alongside the flow. Following informal listening, we observed that for both of these 
approaches fixing $\sigma_{prior}=1$ for the model prior, alongside the predicted context-dependent $\mu_{prior}$, led to better convergence of the model for VC. 

\vspace{-4mm}
\subsubsection{Pre-trained phoneme prior}
\label{subsubsec:edp_prior}
\vspace{-2mm}

\begin{figure}[htb]
  \centering
  \includegraphics[width=8.5cm]{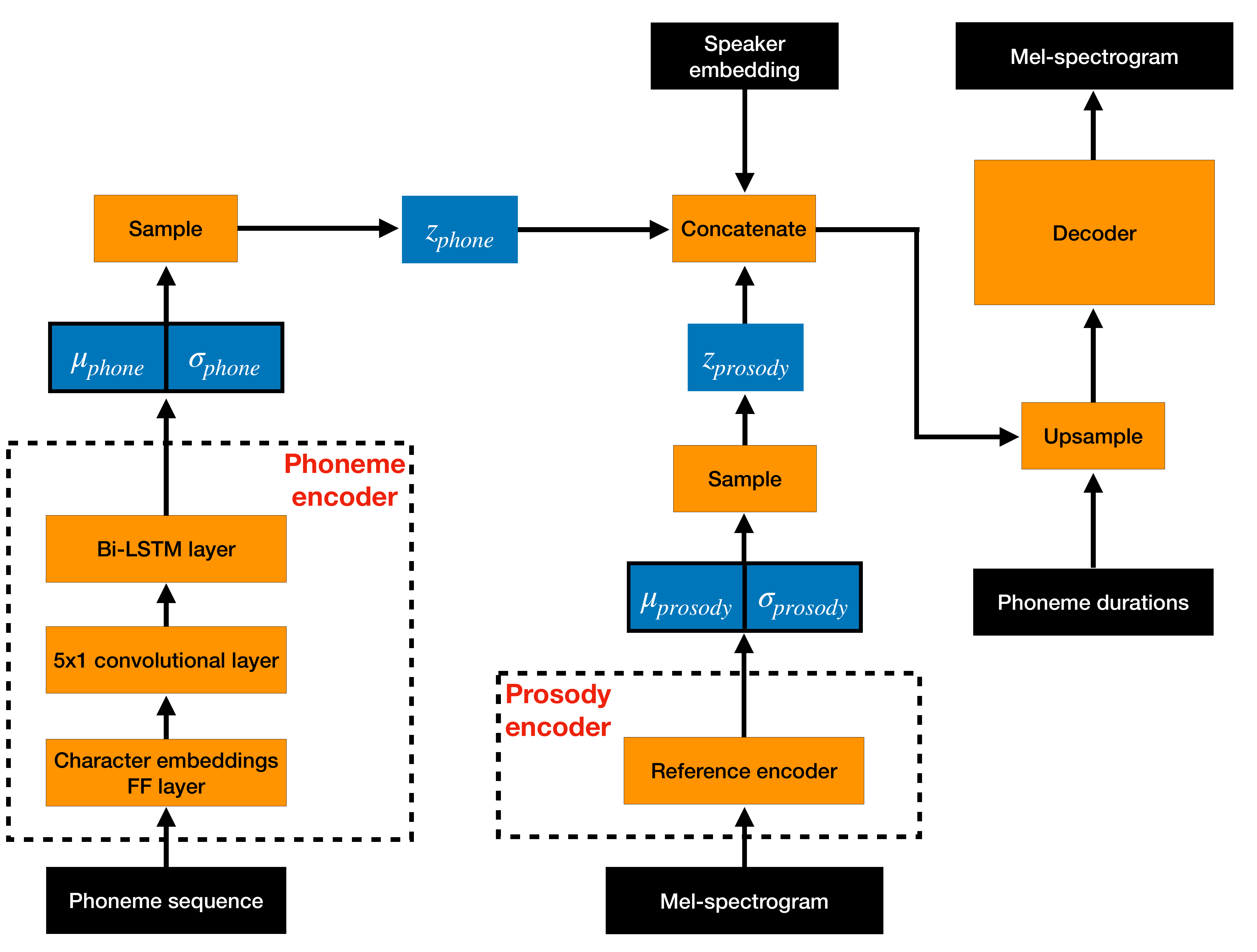}
  \vspace{-4mm}
\caption{Overview of model used to pre-train context-dependent prior for flow-based encoder. $\mu_{phone}$ is used for the prior.}
\label{fig:edp_prior}
\vspace{-4mm}
\end{figure}

The contextual phoneme prior is pre-trained using an attention-free TTS model. This is based on the model presented in \cite{shah2021non} and is shown in Figure \ref{fig:edp_prior}. The only change to this topology from the previously published work is that the phoneme encoder output was adapted to be a variational encoder. Therefore, the vector output by the phoneme encoder is of size 160 (80-dimensional mean vector plus 80-dimensional log standard deviation vector). 
The training schedule to apply a phoneme encoder KLD loss during training matches that used to train the prosody encoder VAE (see \cite{shah2021non} for further details). 
For this investigation $\mu_{phone}$ from this pre-trained model is used as the prior mean ($\mu_{prior}$) to train the flow.

\vspace{-4mm}
\subsubsection{Jointly-trained phoneme prior}
\label{subsubsec:glow_prior}
\vspace{-2mm}
This method 
is closest to the Glow-TTS and SC-Glow-TTS approaches \cite{kim2020glowtts,casanova2021sc}. Here, the flow-based speech encoder 
is trained together with the phoneme encoder that determines the mean of the prior ($\mu_{prior}$). 
The structure of the phoneme encoder is shown in Figure \ref{fig:phone_cnd} ($\mu_{prior}=\mu_{phone}$). 
There is no additional loss that constrains the phoneme encoder output in any way, instead relying fully on Equation \ref{eq:likelihood_loss}.

\vspace{-4mm}
\section{Experimental protocol}
\label{sec:experiments}
\vspace{-4mm}

\subsection{Data}
\vspace{-2mm}
English recordings of approximately 597k utterances from 3173 speakers were used for training. These came from English speakers from 6 different regional accents: US, GB, Canadian, Welsh, Australian, Indian. The amount of data per speaker ranges from 100 to 20k utterances. 
Despite the large range, even speakers with the largest amounts of data still only account for around 3\% of total data, meaning they do not dominate the dataset. 
The data also features a range of different recording conditions with some speakers recorded in studio-quality conditions whilst other speakers were recorded in more ambient surroundings using lower quality microphones. A sampling rate of 16kHz was used for all recordings, from which 80 dimensional mel-spectrograms were extracted using a frame shift of 12.5 ms. For generating audio samples for evaluation, a universal neural vocoder was used \cite{jiao2021universal}.

To evaluate VC quality we take 6 speakers from the training data. These are: US male, US female, Australian female, Indian male, Canadian female and Welsh male. We evaluate all possible combinations of conversion between these speakers, resulting in 30 conversion combinations. We convert 20 utterances from each speaker to each of the other target speakers, resulting in 600 evaluation screens. For accent similarity evaluations we removed comparisons where the source and target speaker are from the same accent (i.e., US male to US female and vice-versa), resulting in 560 evaluation screens.

\vspace{-4mm}
\subsection{Systems}
\vspace{-2mm}

The flow-based VC approaches investigated are: 1) $flow$ $text$-$conditioned$ (Section \ref{subsec:txt_cnd}), 2) $flow$ $text$-$free$ $pre$-$train$ (Section \ref{subsubsec:edp_prior}), 3) $flow$ $text$-$free$ $joint$-$train$ (Section \ref{subsubsec:glow_prior}). 
These models consist of 12 flow steps. The squeeze operator was set to 2. After every 4 flow steps the vector dimensionality is reduced by 16 via the split operation. 

We baseline the performance of the flow-based VC models against state-of-the-art text-conditioned and text-free VC approaches. For the text-conditioned baseline we use CopyCat \cite{karlapati2020copycat}. We applied the small modification to the CopyCat approach introduced in \cite{huybrechts2021low} where source speaker embeddings are concatenated to the upsampled phonemes before passing to the phoneme encoder. This was found to reduce occurrences of speaker leakage. For the text-free baseline we use AutoVC \cite{qian2019autovc}. We found that for this system careful tuning of the compression across time and information bottleneck size of the content encoder was required. Time-wise compression of 16 frames and a bottleneck size in the content encoder of 8 was selected 
following informal listening.

All evaluated systems used the same pre-trained speaker embeddings. In addition, the same phoneme alignments are used to label the data for all systems. Separate front-ends were used per accent. The one-hot phone-set used for all models is the superset of phonemes returned by these different front-ends. Text-conditioned models used an accent embedding to allow the model to disambiguate phonemes according to the accent being used.

\vspace{-2mm}
\subsection{Evaluation}
\vspace{-2mm}
Subjective evaluations of the models were conducted via MUSHRA tests \cite{bureau2003method} using the following metrics:
\begin{itemize}
    \item \textbf{Naturalness:} \textit{``Please rate the audio samples in terms of their naturalness''}. The recording from the target speaker is included among the systems to be rated as an upper-anchor.
    \item \textbf{Speaker similarity:} \textit{``Please listen to the speaker in the reference sample first. Then rate how similar the speakers in each system sound compared to the reference speaker''}. Two different recordings from the target speaker are included. One as the reference sample and the other as one of the systems to be rated, as an upper-anchor. In addition, the source speech recording is included among the systems to be rated, as a lower-anchor.
    \item \textbf{Accent similarity:} \textit{``Please listen to the speaker's accent in the reference sample first. Then rate how similar the accents in each system sound compared to the reference accent''}. Recordings from two different non-source speakers with the source accent are included. One is used for the reference sample and the other is one of the systems to be rated as an upper-anchor. In addition, a non-target speaker with the target speaker's accent is included among the systems to be rated as a lower-anchor. These evaluations ensure that speaker conversion is achieved without changing the source speaker's accent to that of the target speaker (i.e., speaker identity is successfully disentangled from accent).
\end{itemize}

Statistically significant differences between systems were detected using paired t-tests with Holm-Bonferroni correction applied. 
All reported significant differences are for $p \leq 0.05$.

\vspace{-2mm}
\section{Experimental results}
\vspace{-2mm}
\subsection{Word Error Rate analysis}
\label{subsec:wer}
\vspace{-2mm}

To measure the intelligibility of converted speech from the range of models we are investigating, we conduct Word Error Rate (WER) analysis. The AWS Transcribe ASR system was used to transcribe converted speech. The WER is then computed between the sentence text and the transcription. Here we present the WER computed across all 600 test utterances. The US-English ASR model is used to transcribe all utterances. This can introduce some false errors due to the range of English accents being investigated here, however numbers are comparable across different systems.

WER results are shown in Table \ref{tab:wer}. AutoVC is the obvious outlier here, suffering from significantly increased numbers of intelligibility issues following VC. All other systems have similar performance in terms of intelligibility. This includes the text-free VC approaches using normalised flows, a noticeable improvement over the state-of-the-art.

\begin{table}[h!]
\begin{center}
\scalebox{0.83}{
\begin{tabular}{|
>{\columncolor[HTML]{EFEFEF}}l |c|
>{\columncolor[HTML]{EFEFEF}}l |c|
>{\columncolor[HTML]{EFEFEF}}l |c|
}
\hline
\textbf{System} & \textbf{WER} \\ \hline
\textit{flow text-conditioned} & 12.65\% $\pm$ 0.78 \\
\textit{flow text-free pre-train} & 12.38\% $\pm$ 0.77 \\
\textit{flow text-free joint-train} & 12.75\% $\pm$ 0.78 \\
\textit{CopyCat (text-conditioned)} & 11.52\% $\pm$ 0.75\\
\textit{AutoVC (text-free)} & 28.97\% $\pm$ 1.06\\
\hline
\end{tabular}}
\end{center}
\vspace{-6mm}
\caption{\footnotesize{Word Error Rate (WER) with 95\% confidence intervals computed across the test set for all combinations of test speaker conversions.}}
\vspace{-5mm}
\label{tab:wer}
\end{table}

\vspace{-2mm}
\subsection{Comparison of flow-based approaches}
\label{subsec:floweval}
\vspace{-2mm}

Results from naturalness and speaker similarity evaluations are shown in Table \ref{tab:flow_compare}. Differences between all systems are found to be statistically significant except between $flow$ $text$-$conditioned$ and $flow$ $text$-$free$ $pre$-$train$. This demonstrates that for VC with flow-based models we do not observe improvements from text conditioning. This is in contrast to VC literature where typically text conditioning is used to achieve improvements 
over 
text-free VC (for example CopyCat's improvements over AutoVC come from such conditioning). In addition, these results demonstrate the challenge of jointly-training the normalising flow's prior alongside the weights of the flow. By breaking the problem down into pre-training of the prior and then training the flow to maximise the likelihood of this prior, we are able to achieve a significantly better VC model. This is in contrast to Glow-TTS approaches from literature where typically the prior is jointly-trained \cite{kim2020glowtts,casanova2021sc}. 
This indicates that future work to address the challenges of training normalising flows with moving priors may bring improvements to not only VC, but also the Glow-TTS approach. 
Exploring pre-trained priors (work started here) further or investigating improved training schedules, are of interest for such future work.

\begin{table}[h!]
\begin{center}
\scalebox{0.83}{
\begin{tabular}{|
>{\columncolor[HTML]{EFEFEF}}l|c|c|
}
\hline
\textbf{System} & \textbf{Mean rating} & \textbf{Median rating} \\ \hline
\multicolumn{3}{|c|}{\cellcolor[HTML]{C0C0C0}Naturalness} \\ \hline
\textit{target speaker recording} & 72.36 & 76 \\
\textit{flow text-conditioned} & 70.88 & 74 \\
\textit{flow text-free pre-train} & 70.80 & 75 \\
\textit{flow text-free joint-train} & 70.15 & 74 \\
\hline
\multicolumn{3}{|c|}{\cellcolor[HTML]{C0C0C0}Speaker similarity} \\ \hline
\textit{target speaker recording} & 74.98 & 79 \\
\textit{flow text-conditioned} & 64.47 & 72 \\
\textit{flow text-free pre-train} & 64.48 & 73 \\
\textit{flow text-free joint-train} & 62.84 & 71 \\
\textit{source speaker recording} & 60.06 & 71 \\
\hline
\end{tabular}}
\end{center}
\vspace{-6mm}
\caption{\footnotesize{Mean and median MUSHRA scores for naturalness and speaker similarity evaluations comparing different flow-based VC approaches.}}
\label{tab:flow_compare}
\vspace{-4mm}
\end{table}

\vspace{-4mm}
\subsection{Comparison to baseline VC approaches}
\label{subsec:txtcnd}
\vspace{-2mm}

Results from naturalness, speaker similarity and accent similarity evaluations are shown in Table \ref{tab:baseline_vc_eval}. Differences between all systems in naturalness, speaker similarity and accent similarity evaluations are statistically significant except 
between $CopyCat$ and $flow$ $text$-$free$ $pre$-$train$ in the accent similarity evaluation. These results demonstrate that text-free VC with normalising flows outperforms the state-of-the-art for text-free VC, however is currently below the state-of-the-art for text-conditioned VC. 

\begin{table}[h!]
\begin{center}
\scalebox{0.83}{
\begin{tabular}{|
>{\columncolor[HTML]{EFEFEF}}l|c|c|
}
\hline
\textbf{System} & \textbf{Mean rating} & \textbf{Median rating} \\ \hline
\multicolumn{3}{|c|}{\cellcolor[HTML]{C0C0C0}Naturalness} \\ \hline
\textit{target speaker recording} & 69.59 & 73 \\
\textit{flow text-free pre-train} & 63.17 & 66 \\
\textit{CopyCat (text-conditioned)} & 64.74 & 68 \\
\textit{AutoVC (text-free)} & 60.93 & 64 \\
\hline
\multicolumn{3}{|c|}{\cellcolor[HTML]{C0C0C0}Speaker similarity} \\ \hline
\textit{target speaker recording} & 74.57 & 80 \\
\textit{flow text-free pre-train} & 54.01 & 55 \\
\textit{CopyCat (text-conditioned)} & 55.07 & 56 \\
\textit{AutoVC (text-free)} & 52.79 & 50 \\
\textit{source speaker recording} & 47.39 & 50 \\
\hline
\multicolumn{3}{|c|}{\cellcolor[HTML]{C0C0C0}Accent similarity} \\ \hline
\textit{target speaker accent} & 50.70 & 50  \\
\textit{flow text-free pre-train} & 62.47 & 65 \\
\textit{CopyCat (text-conditioned)} & 62.04 & 65 \\
\textit{AutoVC (text-free)} & 59.94 & 63 \\
\textit{source speaker accent} & 68.94 & 74 \\
\hline
\end{tabular}}
\end{center}
\vspace{-6mm}
\caption{\footnotesize{Mean and median MUSHRA scores for naturalness, speaker similarity and accent similarity evaluations comparing text-free flow-based VC approach against baseline systems.}}
\vspace{-6mm}
\label{tab:baseline_vc_eval}
\end{table}

Further analysis was conducted on the MUSHRA responses to look at listener `preferences' between $flow$ $text$-$conditioned$ and $CopyCat$ 
(i.e., for each MUSHRA screen the scores awarded to the systems are interpreted as a preference for one system over the other). These preference scores are then tested for statistical significance using a binomial test, with Bonferonni correction applied. This analysis found a statistically significant preference for $CopyCat$ in terms of naturalness, however the preferences for speaker similarity and accent similarity between these two systems are not significant. Following informal listening we believe that listeners are attending to a larger presence of light audio artefacts introduced by the flow model compared to the $CopyCat$ model. As the analysis 
in Section \ref{subsec:wer} shows, these artefacts are not affecting intelligibility but 
result in a preference for the $CopyCat$ approach. However, in terms of speaker similarity it appears as though preferences between the two approaches are much closer with the magnitude being larger in cases where $CopyCat$ is preferred.
These analyses indicate that whilst normalising flows are able to achieve quality parity between text-conditioned and text-free conversion approaches (within normalising flow-based VC approaches) there is still work required to reduce artefacts.

The MUSHRA screen `preference' analysis was repeated between $AutoVC$ and $flow$ $text$-$free$ $pre$-$train$. This found significant preferences for $flow$ $text$-$free$ $pre$-$train$ over $AutoVC$ for naturalness and accent similarity, however preferences for speaker similarity between these two systems are not significant. This demonstrates that the improvements from the normalising flow over $AutoVC$ 
are focused around perceived naturalness and the 
flow being better equipped to address the problem of disentanglement, 
compared to the more naive compression-based approach of AutoVC. The flow was able to preserve the speaker's accent due to 
its lossless representation, whereas with AutoVC the model appears to have struggled more to disentangle phonemes from speaker identity.

\vspace{-4mm}
\section{Conclusion}
\label{sec:conclusion}
\vspace{-2mm}

In this investigation we explored the use of normalising flows for VC. 
Flows  
learn a lossless representation of speech, in contrast to typical VC models from literature. 
The investigation found that normalising flows experience no degradation in quality from not having prior knowledge of the text being read, allowing for many more applications of VC technology (for example dynamic voice filtering for entertainment purposes). 
The proposed approach successfully disentangles speaker identity from speaker accent, allowing for VC without losing the accent of the source speaker. 
We have demonstrated that the flow-based VC approach brings significant improvements over 
state-of-the-art 
text-free VC, however work is still required to reach the state-of-the-art in text-conditioned models. 
We hypothesise this gap is due to a lack of recurrent connections in the flow decoder, which allow to 
remove artefacts across time. Investigating the use of auto-regressive flows to provide across-frame awareness is left as future work. 
Finally, we demonstrated that joint-training of the flow's prior with the flow weights, the typical approach in Glow-TTS literature, is not optimal and further work into improved training schedules 
is required. 
Here we investigated VC with the flow-based models however this framework allows for independent control of multiple speech attributes (for example control of accent and prosody). Investigating these controls is left as future work.

\pagebreak

\bibliographystyle{IEEEbib}
\bibliography{refs}

\begin{thebibliography}{10}

\bibitem{mohammadi2017overview}
Seyed~Hamidreza Mohammadi and Alexander Kain,
\newblock ``An overview of voice conversion systems,''
\newblock {\em Speech Communication}, vol. 88, 2017.

\bibitem{sisman2020overview}
Berrak Sisman, Junichi Yamagishi, Simon King, and Haizhou Li,
\newblock ``An overview of voice conversion and its challenges: From
  statistical modeling to deep learning,''
\newblock {\em IEEE/ACM Transactions on Audio, Speech, and Language
  Processing}, 2020.

\bibitem{lorenzo2018voice}
Jaime Lorenzo-Trueba, Junichi Yamagishi, Tomoki Toda, Daisuke Saito, Fernando
  Villavicencio, Tomi Kinnunen, and Zhenhua Ling,
\newblock ``The voice conversion challenge 2018: Promoting development of
  parallel and nonparallel methods,''
\newblock in {\em The Speaker and Language Recognition Workshop}, 2018.

\bibitem{qian2019autovc}
Kaizhi Qian, Yang Zhang, Shiyu Chang, Xuesong Yang, and Mark Hasegawa-Johnson,
\newblock ``{AutoVC: Zero-shot voice style transfer with only autoencoder
  loss},''
\newblock in {\em ICML}, 2019.

\bibitem{karlapati2020copycat}
Sri Karlapati, Alexis Moinet, Arnaud Joly, Viacheslav Klimkov, Daniel
  S{\'a}ez-Trigueros, and Thomas Drugman,
\newblock ``{CopyCat: Many-to-many fine-grained prosody transfer for neural
  text-to-speech},''
\newblock in {\em Interspeech}, 2020.

\bibitem{van2017neural}
Aaron van~den Oord, Oriol Vinyals, and Koray Kavukcuoglu,
\newblock ``Neural discrete representation learning,''
\newblock in {\em Advances in Neural Information Processing Systems}, 2017.

\bibitem{polyak2020tts}
Adam Polyak, Lior Wolf, and Yaniv Taigman,
\newblock ``{TTS Skins: Speaker Conversion via ASR},''
\newblock in {\em Interspeech}, 2020.

\bibitem{qian2020unsupervised}
Kaizhi Qian, Yang Zhang, Shiyu Chang, Mark Hasegawa-Johnson, and David Cox,
\newblock ``Unsupervised speech decomposition via triple information
  bottleneck,''
\newblock in {\em ICML}, 2020.

\bibitem{wang2021adversarially}
Jie Wang, Jingbei Li, Xintao Zhao, Zhiyong Wu, and Helen Meng,
\newblock ``Adversarially learning disentangled speech representations for
  robust multi-factor voice conversion,''
\newblock in {\em Interspeech}, 2021.

\bibitem{kim2020glowtts}
Jaehyeon Kim, Sungwon Kim, Jungil Kong, and Sungroh Yoon,
\newblock ``{Glow-TTS: A Generative Flow for Text-to-Speech via Monotonic
  Alignment Search},''
\newblock in {\em Advances in Neural Information Processing Systems}, 2020.

\bibitem{miao2020flow}
Chenfeng Miao, Shuang Liang, Minchuan Chen, Jun Ma, Shaojun Wang, and Jing
  Xiao,
\newblock ``{Flow-TTS: A non-autoregressive network for text to speech based on
  flow},''
\newblock in {\em ICASSP}, 2020.

\bibitem{valle2020flowtron}
Rafael Valle, Kevin Shih, Ryan Prenger, and Bryan Catanzaro,
\newblock ``Flowtron: an autoregressive flow-based generative network for
  text-to-speech synthesis,''
\newblock in {\em ICLR}, 2021.

\bibitem{casanova2021sc}
Edresson Casanova, Christopher Shulby, Eren G{\"o}lge, Nicolas~Michael
  M{\"u}ller, Frederico~Santos de~Oliveira, Arnaldo~Candido Junior, Anderson
  da~Silva Soares, Sandra~Maria Aluisio, and Moacir~Antonelli Ponti,
\newblock ``{SC-GlowTTS: an Efficient Zero-Shot Multi-Speaker Text-To-Speech
  Model},''
\newblock in {\em Interspeech}, 2021.

\bibitem{oord2018parallel}
Aaron Oord, Yazhe Li, Igor Babuschkin, Karen Simonyan, Oriol Vinyals, Koray
  Kavukcuoglu, George Driessche, Edward Lockhart, Luis Cobo, Florian Stimberg,
  et~al.,
\newblock ``Parallel wavenet: Fast high-fidelity speech synthesis,''
\newblock in {\em ICML}, 2018.

\bibitem{prenger2019waveglow}
Ryan Prenger, Rafael Valle, and Bryan Catanzaro,
\newblock ``Waveglow: A flow-based generative network for speech synthesis,''
\newblock in {\em ICASSP}, 2019.

\bibitem{serra2019blow}
Joan Serr{\`a}, Santiago Pascual, and Carlos Segura,
\newblock ``Blow: a single-scale hyperconditioned flow for non-parallel
  raw-audio voice conversion,''
\newblock in {\em Advances in Neural Information Processing Systems}, 2019.

\bibitem{kingma2018glow}
Durk~P Kingma and Prafulla Dhariwal,
\newblock ``Glow: Generative flow with invertible 1x1 convolutions,''
\newblock {\em Advances in Neural Information Processing Systems}, 2018.

\bibitem{ho2019flow++}
Jonathan Ho, Xi~Chen, Aravind Srinivas, Yan Duan, and Pieter Abbeel,
\newblock ``Flow++: Improving flow-based generative models with variational
  dequantization and architecture design,''
\newblock in {\em ICML}, 2019.

\bibitem{kobyzev2020normalizing}
Ivan Kobyzev, Simon Prince, and Marcus Brubaker,
\newblock ``Normalizing flows: An introduction and review of current methods,''
\newblock {\em IEEE Transactions on Pattern Analysis and Machine Intelligence},
  2020.

\bibitem{shen2018natural}
Jonathan Shen, Ruoming Pang, Ron~J Weiss, Mike Schuster, Navdeep Jaitly,
  Zongheng Yang, Zhifeng Chen, Yu~Zhang, Yuxuan Wang, Rj~Skerrv-Ryan,
  Saurous~Rif A., Yannis Agiomyrgiannakis, and Yonghui Wu,
\newblock ``Natural tts synthesis by conditioning wavenet on mel spectrogram
  predictions,''
\newblock in {\em ICASSP}, 2018.

\bibitem{spk_acc_embeddings}
Li~Wan, Quan Wang, Alan Papir, and Ignacio~Lopez Moreno,
\newblock ``{Generalized End-to-End Loss for Speaker Verification},''
\newblock in {\em {ICASSP}}, 2018.

\bibitem{shah2021non}
Raahil Shah, Kamil Pokora, Abdelhamid Ezzerg, Viacheslav Klimkov, Goeric
  Huybrechts, Bartosz Putrycz, Daniel Korzekwa, and Thomas Merritt,
\newblock ``{Non-Autoregressive TTS with Explicit Duration Modelling for
  Low-Resource Highly Expressive Speech},''
\newblock in {\em Speech Synthesis Workshop}, 2021.

\bibitem{jiao2021universal}
Yunlong Jiao, Adam Gabry{\'s}, Georgi Tinchev, Bartosz Putrycz, Daniel
  Korzekwa, and Viacheslav Klimkov,
\newblock ``Universal neural vocoding with parallel wavenet,''
\newblock in {\em ICASSP}, 2021.

\bibitem{huybrechts2021low}
Goeric Huybrechts, Thomas Merritt, Giulia Comini, Bartek Perz, Raahil Shah, and
  Jaime Lorenzo-Trueba,
\newblock ``Low-resource expressive text-to-speech using data augmentation,''
\newblock in {\em ICASSP}, 2021.

\bibitem{bureau2003method}
International Telecommunication Union~Radiocommunication Assembley,
\newblock ``Method for the subjective assessment of intermediate quality level
  of coding systems,''
\newblock {\em Recommendation ITU-R BS. 1534-1}, 2003.

\end{thebibliography}

\end{document}